\documentclass{preprint}
\usepackage{times}
\usepackage{mhequ}
\usepackage{mhenvs}
\usepackage{mhfig}
\usepackage{figurelist}
\usepackage{ScriptFonts}
\usepackage{Symbols}

\def\poly#1#2{{{\bf Pol}_{#1}^{#2}}}
\def\Schwartz{{\cscr S}_{\!n}}
\def\Bound{{\cscr B}}
\def\weight{{\bar\Lambda}}
\def\Prob{{\bf P}}

\definecolor{martin}{rgb}{1,0.2,0.7}
\definecolor{JP}{rgb}{0,0.75,0}

\def\rL{{r_{\!L}}}
\def\rR{{r_{\!R}}}
\def\V#1{V_{#1}}

\newtheorem{assumption}{Assumption}
\newtheorem{notation}[lemma]{Notation}

\ifx\pdfoutput\undefined
\else
	\input protcode.tex
\fi

\def\ppppp{non-degenerate}

\def\scalb{\sml(1+\|x\|^2\smr)}
\def\CK{{\cal K}}
\def\CKK{{\cal L}}

\begin{document}

\ifx\pdfoutput\undefined
\else
    \setprotcode\font
    {\it \setprotcode\font}
    {\bf \setprotcode\font}
    {\bf \it \setprotcode\font}
    \pdfprotrudechars=1
\fi

\date{}
\title{Spectral Properties of Hypoelliptic Operators}
\author{J.-P.~Eckmann\inst{1}\fnmsep\inst{2} and M.~Hairer\inst{1}}
\institute{D\'epartement de Physique Th\'eorique, Universit\'e de Gen\`eve \and
Section de Math\'ematiques, Universit\'e de Gen\`eve
 \\ \email{Jean-Pierre.Eckmann@physics.unige.ch}\\
\email{Martin.Hairer@physics.unige.ch}}
\titleindent=0.65cm

\maketitle
\thispagestyle{empty}
\begin{abstract}
\ifx\pdfoutput\undefined
\else
    \setprotcode\font
    {\it \setprotcode\font}
    {\bf \setprotcode\font}
    {\bf \it \setprotcode\font}
    \pdfprotrudechars=1
\fi
We study hypoelliptic operators with
polynomially bounded coefficients that are of the form
$
K = \sum_{i=1}^m X_i^T X_i^{} + X_0 + f
$,
where the $X_j$ denote first order differential operators, $f$ is a
function with at most polynomial growth, and $X_i^T$ denotes the formal adjoint
of $X_i$ in $\L^2$. For any $\eps>0$ we show that an inequality of the form $ 
\|u\|_{\delta,\delta} \le C\(\|u\|_{0,\eps} + \|(K+iy) u\|_{0,0}\)$
holds for suitable $\delta $ and $C$ which are independent of $y\in\R$,
in weighted Sobolev spaces (the first
index is the derivative, and the second the growth). We apply this
result to the Fokker-Planck operator for an anharmonic chain of
oscillators coupled
to two heat baths. Using a method of
H\'erau and Nier \cite{HN02}, we conclude that its spectrum lies in a
cusp $\{x+iy~|~ x\ge |y|^\tau-c, \tau\in(0,1],c\in\R \}$.
\end{abstract}
\section{Introduction}

In an interesting paper, \cite{HN02}, H\'erau and Nier studied the
Fokker-Planck equation associated to a Hamiltonian system $H$ 
in contact with a heat reservoir at inverse temperature $\beta$.
For this problem, it is
well-known that the Gibbs measure
\begin{equ}
\mu_\beta(dp\,dq) = \exp \( - \beta H(p,q)\)\,dp\,dq 
\end{equ}
is the only invariant measure for the system. 
In their study of convergence under the flow of any measure to the
invariant measure, they were
led to study spectral properties of the Fokker-Planck operator
$\CL$
when considered as an operator on $\L^2(\mu_\beta)$. In particular,
they showed that $\CL$ has a compact resolvent and that its spectrum is
located in a cusp-shaped 
region, as depicted in Figure~\ref{fig:cusp} below, improving (for a
special case) earlier results obtained by Rey-Bellet and Thomas
\cite{LTH}, who showed that $e^{-\CL t}$ is compact and that $\CL$ has spectrum
only in $\rm Re \lambda >c>0$ aside from a simple eigenvalue at 0.
\begin{figure}[h]
\begin{center}
\mhpastefig[3/4]{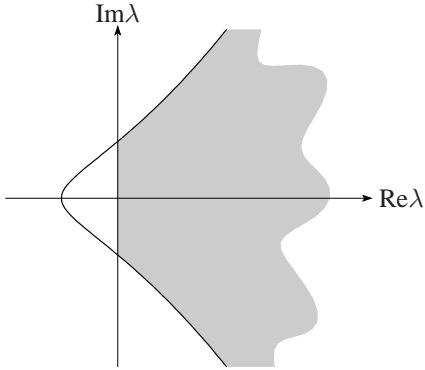}
\end{center}
\caption{Cusp containing the spectrum of $\CL$.}\label{fig:cusp}
\end{figure}

Extending the methods of \cite{HN02}, we show in this paper that the
cusp-shape of the spectrum of $\CL$ occurs for many
H\"ormander-type operators of the form
\begin{equ}[e:K]
K = \sum_{i=1}^m X_i^T X_i^{} + X_0 + f\;,
\end{equ}
(the symbol $^T$ denotes the formal adjoint in $\L^2$) when the family of
vector fields $\{X_j\}_{j=0}^m$ is sufficiently non-degenerate (see
Definition~\ref{def:nondegen} and assumption $b_1$ below) and some growth
condition on $f$ holds.

The main motivation for our paper comes from the study of the 
model of heat conduction proposed in \cite{EPR} and further studied
in \cite{EPR2,EH,LTh00,LTH,LTh02}. These papers deal with Hamiltonian
anharmonic chains of 
point-like particles with nearest-neighbor interactions whose ends are
coupled to heat reservoirs modeled by linear classical field
theories. Our results improve the detailed knowledge about the
spectrum of the generator $\CL$
of the associated Markov process, see Sect.~\ref{sect:example}. 
As a by-product, our paper also gives a more elegant analytic proof of
the results 
obtained in \cite{EH}. A short probabilistic proof has already been obtained
in \cite{LTH}.

The main technical result needed to establish the cusp-form of the
spectrum is the
Sobolev estimate 
\theo{theo:estglob} which seems to be new.

\begin{acknowledge}We thank G. van Baalen and E. Zabey for helpful
remarks.
This work was partially supported by the Fonds
National Suisse.
\end{acknowledge}

\section{Setup and Notations}

We will derive lower bounds for hypoelliptic operators with
polynomially bounded coefficients that are of the form \eref{e:K}. We
start by defining the class of functions and vector fields we consider.
\subsection{Notations}

For $N\in \R$, we define the set $\poly{0}{N}$ of polynomially growing
functions by
\begin{equ}[e:defPoly]
\poly{0}{N} = \Bigl\{f\in \CC^\infty(\R^n) \;\Big|\;\forall
\alpha,\; \sup_{x \in \R^n} \(1+ \|x\|\)^{-N}\d^\alpha f(x) \le
C_\alpha\Bigr\}\;.
\end{equ}
In this expression, $\alpha$ denotes a multi-index of arbitrary order. We also
define the set $\poly{1}{N}$ of vector fields in $\R^n$ that can be written as
\begin{equ}
G\,=\,G_0(x)+\sum_{j=1}^n G_j(x)\d_j\;,\qquad G_i \in \poly{0}{N}\;.
\end{equ}
One can similarly define sets $\poly{k}{N}$ of $k$th order differential
operators. It is clear that if $X \in \poly{k}{N}$ and $Y \in \poly{\ell}{M}$,
then $[X,Y] \in \poly{k+\ell-1}{N+M}$. If $f$ is in $\poly{0}{N}$, but
not in $\poly{0}{N+\eps }$ for any $\eps >0$, we say it is of
degree $N$.
\subsection{Hypotheses}\label{sec:hypotheses}

\begin{definition}\label{def:nondegen}
A family  $\{A_i\}_{i=1}^m $  of vector fields in $\R^n$ with
 $A_i\,=\,\sum_{j=1}^nA_{i,j}\partial _j$ is called {\em
 \ppppp} if there exist constants $N$ and $C$ such that for
every $x \in \R^n$ and every vector $v \in \R^n$ one has the bound
\begin{equ}
\|v\|^2 \le C\(1 + \|x\|^2\)^N \sum_{i=1}^m \scal{A_i(x),v}^2 ~,
\end{equ}
with $\scal{A_i(x),v}=\sum_{j=1}^n A_{i,j}(x)\,v_j$.
\end{definition}

The conditions on $K$ which we will use below are taken from the
following list.
\begin{itemize}
\item[{$a$.}] The vector fields $X_j$ with $j=0,\ldots,m$ belong to
$\poly{1}{N}$ and the function $f$ belongs to $\poly{0}{N}$.
\item[ $b_0$.] There exists a finite number $M$  such that the family
consisting of $\{X_i\}_{i=0}^m$, $\{[X_i,X_j]\}_{i,j=0}^m$,
$\bigl\{\bigl[[X_i,X_j],X_k\bigr]\bigr\}_{i,j,k=0}^m$ and so on up to
commutators of rank $M$ is \ppppp.
\item[$b_1$.] There exists a finite number $M$  such that the family consisting
of $\{X_i\}_{i=1}^m$, $\{[X_i,X_j]\}_{i,j=0}^m$,
$\bigl\{\bigl[[X_i,X_j],X_k\bigr]\bigr\}_{i,j,k=0}^m$ and so on up to
commutators of rank $M$ is \ppppp.
\end{itemize}
The difference between $b_0$ and $b_1$ is in the inclusion of the
vector field $X_0$ (in $b_0$), so that $b_1$ is stronger than $b_0$.

\begin{definition}\label{def:KK}
We call $\CK_0$ the class of operators of
the form of \eref{e:K} satisfying $a$ and $b_0$ above, and $\CK_1$ the
class of those satisfying $a$ and $b_1$. Clearly, $b_1$ is more
restrictive than $b_0$ and therefore $\CK_1\subset \CK_0$.
\end{definition}

\begin{remark}If $K$ is in $\CK_0$ then $K$ is hypoelliptic. If $K$ is
in $\CK_1$ then $\d_t +K$ is hypoelliptic.
\end{remark}

\section{Localized Bound}
\label{sect:local}

%
%
The main result of this section is \theo{theo:loc} which provides
bounds for localized test functions.

We let $\CB(x)$ denote the unit cube around $x\in\R^n$:
\begin{equ}
\CB(x) = \bigl \{y \in\R^n \,~\big|~\, |y_j - x_j| \le 1 \;,\;
j=1,\ldots,n\bigr \}\;.
\end{equ}
To formulate our bounds, we introduce the operator $\Lambda$, defined as the
 positive square root of
 $\Lambda^2 = 1 - \sum_{i=1}^n \d_i^2 = 1-\Delta$.
Later on, we will also need the multiplication operator $\weight$ defined
 as the positive root of (multiplication by)
$\weight^2 = 1 + \|x\|^2$.
\begin{theorem}\label{theo:loc}
Assume $K\in \CK_1$. Then, there exist positive constants $\eps_*$, $C_*$,
and $N_*$ such that for every $x \in \R^n$ and every
$u\in\CC_0^\infty \bigl(\CB(x)\bigr)$, one has {\bf
uniformly} for $y\in\R^n$:
\begin{equ}[e:locest]
\|\Lambda^{\eps_*} u\| \le  C_* \scalb^{N_*}\|u\| + \|(K+iy) u\|\;.
\end{equ}
If $K$ is in $\CK_0$ (but not in $\CK_1$) the same estimate
holds, but the constant $C_*$ will depend generally on $y$.\footnote{The norms are $\L^2$
norms.}

\end{theorem}

\begin{proof}
The novelty of the bound is in allowing for polynomial growth of the
coefficients of the differential operators. Were it not for this, the
result would be a special case
of H\"ormander's
proof of hypoellipticity of second-order partial differential
operators \cite[Thm.~22.2.1]{Ho}. Since the
coefficients of our differential operators can grow polynomially
we need to work with weighted spaces.

We introduce a family of weighted Sobolev spaces $S^{\alpha,\beta}$
with $\alpha,\beta \in\R$ as the
following subset of tempered distributions $\Schwartz'$ on
$\R^n$:
\begin{equ}
S^{\alpha,\beta} = \{u \in \Schwartz' \,|\, \Lambda^{\alpha}\weight^{\beta}
u\in \L^2(\R^n)\}\;.
\end{equ}
We equip this space with the scalar product
\begin{equ}[e:ab]
\scal{f,g}_{\alpha,\beta} =
\scal{\Lambda^{\alpha}
\weight^{\beta}f,\Lambda^{\alpha}\weight^{\beta}g}_{\L^2}\;,
\end{equ}
writing also $\scal{\cdot,\cdot}_\alpha$ instead of
$\scal{\cdot,\cdot}_{\alpha,0}$. We also use the corresponding norms
$\|\cdot\|_{\alpha ,\beta }$. Note that these spaces are actually a particular
case of the more general class of Sobolev spaces introduced in \cite{BCH}.

The following lemma lists a few properties of the spaces $S^{\alpha,\beta}$
that will be useful in the sequel. We postpone its proof to
Appendix~\ref{app:est}.
%
%
\begin{lemma}\label{lem:est}
Let $\alpha,\beta \in \R$. We have the following:
\begin{itemize}
\item[a.] {\rm Embedding:} For $\alpha' \ge \alpha$ and $\beta' \ge \beta$, the
space $S^{\alpha',\beta'}$ is continuously embedded into $S^{\alpha,\beta}$.
The embedding is compact if and only if both inequalities are strict.

\item[b.] {\rm Scales of spaces:} The operators $\Lambda^\gamma$ and
$\weight^\gamma$ are
bounded from $S^{\alpha,\beta}$ into $S^{\alpha-\gamma,\beta}$ and
$S^{\alpha,\beta-\gamma}$ respectively. If $X \in \poly{k}{N}$ then
$X$ is bounded from $S^{\alpha,\beta}$ into
$S^{\alpha-k,\beta-N}$.

\item[c.] {\rm Polarization:} For every $\gamma,\delta \in \R$, one has the
bound
\begin{equ}
\scal{f,g}_{\alpha,\beta} \le C
\,\,\|f\|_{\alpha',\beta'}\,\,\|g\|_{\alpha'',\beta''}\;,\quad \alpha
'+\alpha ''=2\alpha ,\,\,\beta '+\beta ''=2\beta ~,
\end{equ}
which holds for all $f$ and $g$ belonging to the Schwartz space $\Schwartz$.
The constant $C$ may depend on the indices.

\item[d.] {\rm Commutator:} Let $X \in \poly{k}{N}$ and $Y \in
\poly{k'}{N'}$. For every 
$\gamma \in \R$, $[X,\Lambda^\gamma]$ is bounded from $S^{\alpha,\beta}$ into
$S^{\alpha+1-k-\gamma,\beta-N}$. Similarly, $[X,[Y,\Lambda^\gamma]]$ is bounded from $S^{\alpha,\beta}$ into
$S^{\alpha+2-k-k'-\gamma,\beta-N-N'}$.

\item[e.] {\rm Adjoint:} Let $X \in \poly{k}{N}$ and let $f,g
\in \Schwartz$. Then
\begin{equ}
\scal{f,Xg}_{\alpha,\beta} = \scal{X^T f,g}_{\alpha,\beta} + R(f,g)\;,
\end{equ}
where the  bilinear form $R$ satisfies the bound
\begin{equ}
|R(f,g)| \le C
\|f\|_{\alpha',\beta '}\|g\|_{\alpha'',\beta''}\;,
\end{equ}
with
\begin{equ}[e:aabb]
\alpha '+\alpha ''=2\alpha +k-1,\quad \beta '+\beta '' = 2\beta +N~.
\end{equ}
The constant $C$ may depend on the indices.

\end{itemize}
\end{lemma}

\begin{notation}We write $K_y$ instead of $K+iy$. We also
introduce the notation $\Phi \le
\Bound$ to mean: 
There exist constants $C$ and $N$ independent of $x$
and $y$
such that for all $u\in\CC_0^\infty \bigl(\CB(x)\bigr)$:
\begin{equ}
\Phi \le C \(1+\|x\|\)^N\(\|u\| + \|K_y u\|\)\;.
\end{equ}
\end{notation}

We will show below that
\begin{equ}[e:boundA]
\|A \Lambda^{\eps-1} u\| \le \Bound\;,
\end{equ}
holds for $A$ taking values among all of the vector fields appearing in
{$b_1$} or {$b_0$}. 
Assuming \eref{e:boundA} one completes the proof of \theo{theo:loc} as follows:
Notice that if the collection $\{A_i\}_{i=1}^k $ is non-degenerate, then
\begin{equ}
\|\Lambda u\|^2 \le C_1\(1 +
\|x\|^2\)^{N_1} \sum_{i=1}^k \|A_i u\|^2 \;,\end{equ}
for every $x\in\R^n$ and every $u \in
\CC_0^\infty\(\CB(x)\)$. Therefore, by \eref{e:boundA} we find
\begin{equ}
\|u\|_\eps^2 \,=\,\|\Lambda \Lambda^{\eps-1} u\|\,\le\,C_1\(1 +
\|x\|^2\)^{N_1} \sum_{i=1}^k \|A_i \Lambda^{\eps-1}u\|^2\,\le\,\Bound^2~.
\end{equ}
Polarizing, we obtain:
\begin{equs}
\|u\|_{\eps /2}^2\,&\le\,
\|u\|\,\|u\|_{\eps }\,\le\,C_2\|u\|\scalb^{N_2}\(\|u\| + \|K_y
u\|\)\\
\,&\le\,C_2^2\|u\|^2\scalb^{2N_2}+\(\|u\| + \|K_y
u\|\)^2\\
\,&\le\,\(C_2\|u\|\scalb^{N_2}+\|u\| + \|K_y
u\|\)^2~,
\end{equs}
and hence \eref{e:locest} follows with $\eps _*=\eps/2$,
$N_*=N_2$, and $C_*=C_2+1$. 

It remains to prove \eref{e:boundA}.
\begin{remark}To the end of this proof, we use the symbols $C$ and $N$
to denote 
generic
positive constants which may change from one inequality to the
next.
\end{remark}

By the bound on $[A, \Lambda^{\eps-1}]$ of \lem{lem:est}(d)---and
the fact that $u\in \CC_0^\infty(\CB(x))$ implies $\|u\|_{0,N} \le
C\(1 + \|x\|^2\)^{N/2} \|u\|$ for every $N > 0$---we will have shown
\eref{e:boundA} if we can prove
\begin{equ}[e:boundA']
\|A u\|_{\eps-1} \le  \Bound\;.
\end{equ}
Notice that by \lem{lem:est}(b), the estimate \eref{e:boundA'} yields
\begin{equ}[e:boundAgamma]
\|A u\|_{\eps-1,\gamma}^2 \le  C_{\gamma} \scalb^{\gamma + N}\(\|u\|^2 +
\|K_y u\|^2\)\;,
\end{equ}
for every $\gamma > 0$, $x\in\R^n$, and $u\in \CC_0^\infty(\CB(x))$.

To prove \eref{e:boundA'}, we proceed as follows. First, we verify it
for $A= X_i$ with $i=1,\ldots,m$ (as well as for $A = X_0$ in the case
$\CK_0$). The
remaining bounds are shown by induction. 
The induction step consists in proving that if 
\eref{e:boundA'} holds for some $A\in\poly{1}{N}$
then
\begin{equ}[e:induce]
\|[A,
X_i] u\|_{\eps/8-1} \le  \Bound\ \text{  for }\ i=0,\ldots,m~.
\end{equ}

\noindent\textbf{The first step.}
By the definition of $K$ and the fact that $X_i$ maps $\CC_0^\infty(\CB(x))$
into itself, we see that
\begin{equ}[e:nocheine]
\|X_i u\| \le \Bound~,\quad i=1,\dots,m~,
\end{equ}
that is,
\eref{e:boundA'} holds for $\eps \le 1$ and
$A = X_i$. 

We next show
that it also holds for $A = X_0$ whenever $\eps \le 1/2$. (This will
be the
only place in the proof 
where $C$ depends on $y$, but we need this
estimate only for the case $\CK_0$.) Using \eref{e:K} and
\lem{lem:est}(c), we can write
\begin{equ}
\|X_0 u\|_{-1/2}^2 \le \|X_0 u\|_{-1}\(\|K_y u\| + \|fu\| + |y|\,\|u\|\) +
\sum_{i=1}^m\scal{X_0 u, X_i^TX_i^{} u}_{-1/2}\;.
\end{equ}
Using \lem{lem:est}(b) to estimate $\|X_0 u\|_{-1}$, the first term is bounded
by $\Bound^2$, so it remains to
bound $\scal{X_0 u, X_i^TX_i^{} u}_{-1/2}$. Using this time \lem{lem:est}(e),
(with $\alpha =-\frac{1}{2}$ and $\beta =0$),
we write
\begin{equ}[e:decomp]
\scal{X_0 u, X_i^TX_i^{} u}_{-1/2} = \scal{X_i X_0 u,X_i u}_{-1/2} + R(X_0
u,X_i u)\;,
\end{equ}
where $R(X_0 u,X_i u)$ is bounded by $C \|X_0u\|_{-1}\|X_i u\|$, which in
turn is bounded by $\Bound^2$, using the previous bounds on $\|X_0u\|_{-1}$ and
$\|X_i u\|$. The first term of \eref{e:decomp}
can be written as
\begin{equ}
|\scal{X_i X_0 u,X_i u}_{-1/2}| \le C\|X_i X_0 u\|_{-1}\|X_i u\|\;.
\end{equ}
Since $\|X_i u\| \le \Bound$ by \eref{e:nocheine}, we only need to bound $\|X_i
X_0 u\|_{-1}$ by $\Bound$. This is achieved by writing
\begin{equ}
\|X_i X_0 u\|_{-1} \le \|X_0 X_i u\|_{-1} + \|[X_i,X_0]u\|_{-1} \;.
\end{equ}
The second term is bounded by $\Bound$ using \lem{lem:est}(b). The first term
is also
bounded by $\Bound$ since $\|X_i\, u\|_{0,N} \le C\(1+\|x\|\)^N
\|X_i\,u\|$ and $X_0$ is bounded from
$S^{0,N}$ into $S^{-1,0}$ (for some $N$) by \lem{lem:est}(b).
Therefore, we conclude that 
\begin{equ}[e:nocheine2]
\|X_0 u\|_{-1/2} \le \Bound~,
\end{equ}
where $C$ will in general depend on $y$.

\noindent\textbf{The inductive step.}
Let $A \in \poly{1}{N}$ and assume that \eref{e:boundA'} holds. We show that a
similar estimate (with different
values for $\eps$, $C$, and $N$) then also holds for $B = [A,X_i]$ with
$i=0,\ldots,m$. We distinguish the case $i=0$ from the others.

\noindent\textbf{The case $\mathit{i> 0}$.} We assume that
\eref{e:boundA'} holds and we estimate $\|B u\|_{\eps'-1}$ for some
$\eps '\le 1/2$ to be fixed later. We obtain
\begin{equ}
\|B u\|_{\eps'-1}^2 = \scal{B u, A X_i\,u}_{\eps'-1} - \scal{B u, X_i A
u}_{\eps'-1} = T_1 + T_2\;.
\end{equ}
Both terms $T_1$ and $T_2$ are estimated separately. For $T_1$, we get
from \lem{lem:est}(e):
\begin{equ}
T_1 = -\scal{A B u, X_i\,u}_{\eps'-1} + R(B u, X_i\,u)\;,
\end{equ}
where (since $\eps' \le 1/2$),
\begin{equ}[e:boundR]
|R(B u, X_i\,u)| \le C \(1+\|x\|\)^N \|B u\|_{-1}\|X_i\,u\| \le C \(1+\|x\|\)^N \|u\|
\|X_i\,u\| \le \Bound^2\;.
\end{equ}
The term $\scal{A B u, X_i\,u}_{\eps'-1}$ is written as
\begin{equ}
|\scal{A B u, X_i\,u}_{\eps'-1}| \le \|B A u\|_{2\eps'-2}\|X_i\,u\| + \|[A,B]
u\|_{-1} \|X_i\,u\|\;.
\end{equ}
The second term is bounded by $\Bound^2$ like in \eref{e:boundR}.
The first term is also bounded by $\Bound^2$ by
combining \lem{lem:est}(b) with the induction assumption in its form
\eref{e:boundAgamma} (taking $2\eps' \le \eps$). The estimation of
$T_2$ is very similar: we write again
\begin{equ}[e:T2]
T_2 = -\scal{X_i B u, A u}_{\eps'-1} + R(B u,A u)\;.
\end{equ}
The first term is bounded by $C\|X_i B u\|_{-1} \|A u\|_{2\eps'-1}$. The
second factor of this quantity is boun\-ded by $\Bound$ by the inductive
assumption, while the first factor is
bounded by
\begin{equ}[e:boundX_iB]
\|X_i B u\|_{-1} \le \|B X_i\,u\|_{-1} + \|[B,X_i]u\|_{-1} \le \Bound\;,
\end{equ}
using \lem{lem:est}(b) and the estimate $\|X_i\|_{0,N} \le \Bound$.
The remainder $R$ of \eref{e:T2} is bounded by
\begin{equ}
|R(B u,A u)| \le \|Bu\|_{-1}\|A u\|_{2\eps'-1}\;,
\end{equ}
which is bounded by $\Bound^2$, using
\lem{lem:est}(b) for the first factor and the inductive assumption for the
second.
Combining the estimates on $T_1$ and $T_2$ we get
\begin{equ}
\|Bu\|_{\eps'-1} \le \Bound \quad\text{for}\quad \eps' \le \eps/2\;,
\end{equ}
which is the required estimate.


\noindent\textbf{The case $\mathit{i = 0}$.} To conclude the proof of
\theo{theo:loc}, it remains to bound $\|B u\|_{\eps'-1}$ by $\Bound$. In this
expression, $B = [A,X_0]$ and
$\eps' > 0$ is to be fixed later. We first introduce the operator
\begin{equ}
\tilde K = \sum_{i=1}^m X_i^T X_i^{}\;,
\end{equ}
which is (up to a term of multiplication by a function) equal to the real part of $K_y$, when considered as
an operator on $\L^2$. We can thus write $X_0$ as
\begin{equ}
	X_0 = K - \tilde K + f_1 = \tilde K - K^T + f_2\;,
\end{equ}
for two functions $f_1,f_2 \in \poly{0}{N}$ for some $N$. This allows us to
express $B$ as
\begin{equ}
B = [A,X_0] = A K_y + K_y^T A + [\tilde K,A] - 2 \tilde K A + Af_1 - f_2A\;.
\end{equ}
We write $\|Bu\|_{\eps'-1}^2 = \scal{Bu,[A,X_0]u}_{\eps'-1}$ and we
bound separately by $\Bound^2$ each of the terms that appear in this expression
according to the above decomposition of the commutator.

The two terms containing $f_1$ and $f_2$ are bounded by $\Bound^2$ using the
inductive assumption. We therefore concentrate on the four remaining
terms.

\noindent\textbf{The term $\mathit{{A K_y}}$.} We write this term as
\begin{equ}
\scal{Bu,A K_y u}_{\eps'-1} = -\scal{BAu, K_y u}_{\eps'-1} + \scal{[A,B]u, K_y
u}_{\eps'-1} + R(Bu,K_y u)\;,
\end{equ}
where the two last terms are bounded by $\Bound^2$ using \lem{lem:est}(b,e).
Using assumption \eref{e:boundAgamma} (assuming $\eps' \le \eps/2$) and
\lem{lem:est}(b,c), we also bound the first term by $\Bound^2$.

\noindent\textbf{The term $\mathit{{K_y^T A}}$.} We write this
term as
\begin{equ}
\scal{Bu,K_y^T A}_{\eps'-1} = \scal{K_y Bu,A}_{\eps'-1} +
\scal{\Lambda^{2-2\eps'}[K,\Lambda^{2\eps'-2}]Bu,Au}_{\eps'-1} = T_1 + T_2\;.
\end{equ}
The term $T_1$ is bounded by $\|K_y Bu\|_{-1}\|Au\|_{2\eps'-1}$ by
polarization. The second
factor of this product is bounded by $\Bound$, using the induction
hypothesis and the assumption $\eps' \le \eps/2$. The first factor is bounded
by
\begin{equ}[e:KyB]
\|K_y Bu\|_{-1} \le \|B K_y u\|_{-1} + \|[K,B]u\|_{-1}\;.
\end{equ}
The first term of this sum is obviously bounded by $\Bound$. The second term is
expanded using the explicit form of $K$ as given in \eref{e:K}. The only
``dangerous'' terms appearing in this expansion are those of the form
$\|[X_i^TX_i^{},B]u\|_{-1}$. They are bounded by
\begin{equ}
\|[X_i^TX_i^{},B]u\|_{-1}  \le \|[X_i^T,B]X_i^{}u\|_{-1} +
\|[X_i,B]X_i^{T}u\|_{-1}  + \bigl\|\bigl[X_i^{T},[X_i,B]\bigr]u\bigr\|_{-1}
\;.
\end{equ}
The terms in this sum are bounded individually by $\Bound$, using the
estimates on $\|X_i\,u\|$, together with \lem{lem:est}(b,d). We now turn to the term
$T_2$. We
bound it by
\begin{equ}
|T_2| \le
C\|\Lambda^{2-2\eps'}[K,\Lambda^{2\eps'-2}]Bu\|_{-1}\|Au\|_{2\eps'-1}\;.
\end{equ}
The second factor is bounded by $\Bound$ by the induction
hypothesis, so we focus on the first factor. We again write explicitly $K$ as
in
\eref{e:K} and estimate each term separately. The two terms containing
$X_0$ and $f$ are easily bounded by $\Bound$ using \lem{lem:est}(b,d). We also write
$X_i^T X_i^{} = X_i^2 + Y_i$ with $Y_i \in \poly{1}{N}$ and similarly bound by
$\Bound$ the terms in $Y_i$. The remaining terms are of the type
\begin{equ}
Q_i = \|\Lambda^{2-2\eps'}[X_i^2,\Lambda^{2\eps'-2}]Bu\|_{-1}\;.
\end{equ}
They are bounded by
\begin{equ}
Q_i \le 2\|\Lambda^{2-2\eps'}[X_i,\Lambda^{2\eps'-2}] X_i Bu\|_{-1} +
\bigl\|\Lambda^{2-2\eps'}\bigl[X_i,[X_i,\Lambda^{2\eps'-2}]\bigr]
Bu\bigr\|_{-1}~.
\end{equ}
In order to bound the first term, one writes $X_i B = B X_i + [X_i,B]$ and
bounds each term separately by $\Bound$, using the bound $\|X_i u\|_{0,\gamma}
\le \Bound$ together with \lem{lem:est}(b,d). The last term is also bounded by
$\Bound$ using \lem{lem:est}(d).

\noindent\textbf{The term $\mathit{{[\tilde K,A]}}$.} We write $\tilde K =
\sum_{i=1}^m X_i^T X_i^{}$ and we bound each term separately:
\begin{equs}
\scal{Bu,[X_i^T X_i^{},A] u}_{\eps'-1} &= \scal{Bu,X_i^T [X_i^{},A]
u}_{\eps'-1}
+ \scal{Bu,[X_i^T,A] X_i\,u}_{\eps'-1} \\
	&\equiv T_{i,1} + T_{i,2}\;.
\end{equs}
The first term is written as
\begin{equ}
T_{i,1} = \scal{X_i B u, [X_i^{},A] u}_{\eps'-1}
+ R(u)\;,
\end{equ}
where $R(u)$ is bounded by $C\|B u\|_{-1}\|[X_i^{},A] u\|_{2\eps'-1}$. The
first
factor is bounded by $\Bound$ using \lem{lem:est}(b) and the second factor is
bounded by $\Bound$, using the
estimate for the case $i \neq 0$ (we have to assume $\eps' \le \eps/4$ in order
to get this bound). The term $\scal{X_i B u, [X_i^{},A]
u}_{\eps'-1}$ is estimated by
\begin{equ}
|\scal{X_i B u, [X_i^{},A] u}_{\eps'-1}| \le \|X_i B u\|_{-1} \|[X_i^{},A]
u\|_{2\eps'-1}\;.
\end{equ}
The first factor is bounded by $\Bound$ as in \eref{e:boundX_iB} and the second
factor is
again bounded by $\Bound$, using the estimate for the case $i\neq 0$. It thus
remains to bound
$T_{i,2}$, which we write as
\begin{equ}
T_{i,2} = \scal{Bu, X_i [X_i^T,A] u}_{\eps'-1} +
\scal[b]{Bu,\bigl[[X_i^T,A], X_i\bigr] u}_{\eps'-1}\;.
\end{equ}
The first term in this equation is similar to the term $\scal{Bu,X_i^T
[X_i^{},A] u}_{\eps'-1}$ and is bounded by $\Bound^2$ in the same way. The
second term is
bounded by
\begin{equ}
\scal[b]{Bu,\bigl[[X_i^T,A], X_i\bigr] u}_{\eps'-1} \le \|Bu\|_{-1}
\bigl\|\bigl[[X_i^T,A], X_i\bigr] u\bigr\|_{2\eps'-1}\;,
\end{equ}
which can also be bounded by $\Bound^2$, using the estimate for the case $i
\neq 0$, provided
$\eps' \le \eps/8$.

\noindent\textbf{The term $\mathit{{\tilde K A}}$.} In order to bound this
term, we need the following preliminary lemma:
\begin{lemma}\label{lem:inter}
Let $v \in\Schwartz$, $\alpha,\delta \in\R$, and let $K_y$ be as above. There
exist constants $\tilde C$ and $\tilde N$ independent of $y$ such that the estimate
\begin{equ}[e:estKyscal]
\Bigl|\Re \scal{K_y v,v}_{\alpha} - \sum_{i=1}^m \|X_i v\|_{\alpha}^2 \Bigr|
\le \tilde C\sum_{i=1}^m \|X_i v\|_{\alpha - \delta,\tilde N}
\|v\|_{\alpha+\delta,\tilde N} + \tilde C
\|v\|_{\alpha,\tilde N}^2\;,
\end{equ}
holds.
\end{lemma}
\begin{proof}
Obviously $\scal{K_y v,v}_{\alpha}=\scal{K v,v}_{\alpha}$.
We decompose $K$ according to
\eref{e:K}. The terms containing $X_0$ and $f$ are bounded by
$C \|v\|_{\alpha,N}^2$ according to \lem{lem:est}(b,e), so we focus
on the terms containing $X_i^T X_i^{}$. Using \lem{lem:est}(e), we write them
as
\begin{equ}
\scal{X_i^T X_i^{} v,v}_{\alpha} = \|X_i v\|_{\alpha}^2 + R_i(v)\;,
\end{equ}
where $R_i(v)$ is bounded by $C\|X_i v\|_{\alpha -
\delta,N}\|v\|_{\alpha+\delta,N}$. This concludes the proof of \lem{lem:inter}.
\end{proof}

We now write the term containing $\tilde K A$ as
\begin{equ}[e:besser1]
\scal{Bu,\tilde K Au}_{\eps'-1} = \sum_{i=1}^m \(\scal{X_i Bu,X_i Au}_{\eps'-1} +
R_i\)\;,
\end{equ}
and we apply \lem{lem:est}(e) with $f=Bu$, $g=X_iAu$, $X=X_i^T$. Then
we find
\begin{equ}
|R_i| \le \|Bu\|_{-1,N}\|X_i 
Au\|_{2\eps'-1} \le  \|Bu\|_{-1,N}^2 + \|X_i Au\|_{2\eps'-1}^2\;.
\end{equ}
By \lem{lem:est}(b), the first term is bounded by $\Bound^2$. Using \lem{lem:est}(c) 
to polarize the scalar product in \eref{e:besser1} we thus get
\begin{equ}
|\scal{Bu,\tilde K Au}_{\eps'-1} | \le \Bound^2 + C\sum_{i=1}^m \|X_i
Bu\|_{-1}^2 + C\sum_{i=1}^m \|X_i Au\|_{2\eps'-1}^2\;.
\end{equ}
The term involving $\|X_i Bu\|_{-1}^2$ is bounded by $\Bound^2$ as in \eref{e:boundX_iB}. 
The last term is bounded by \lem{lem:inter}, yielding
\begin{equs}
|\scal{Bu,\tilde K Au}_{\eps'-1} | &\le \Bound^2 + C |\scal{K_y Au,Au}_{2\eps'-1}| + C\sum_{i=1}^m \|X_i A
u\|_{-1,\tilde N}^2 \\
	&\quad + C\|Au\|_{4\eps'-1,\tilde N}^2\;.
\end{equs}
The last term in this expression is bounded by $\Bound^2$ by the induction
hypothesis if
we choose $\eps' \le \eps/4$. The term containing $X_i A u$ can be
bounded by
$\Bound^2$ as in \eref{e:boundX_iB}, so the only term that remains to be bounded is $|\scal{K_y
Au,Au}_{2\eps'-1}|$. By polarizing the estimate obtained by \lem{lem:est}(c),
one gets
\begin{equ}
|\scal{K_y Au,Au}_{2\eps'-1}| \le C\|Au\|_{4\eps'-1}^2 + C\|K_y A
u\|_{-1}^2\;.
\end{equ}
The first term is bounded by $\Bound^2$ using the induction assumption. The
second term is
bounded by $\Bound^2$ exactly like \eref{e:KyB} above. 
Summing all these bounds this proves \eref{e:induce} and hence the inductive step is completed.

Since $K$ was assumed to satisfy $\CK_1$ (or $\CK_0$), we see that
after $M$ inductive steps the 
proof of  \theo{theo:loc} is complete.
\end{proof}

\section{Global Estimate}

The results of the previous section were restricted to functions $u$ with
well-localized compact support. In this section, we are interested in getting
bounds for every $u\in \Schwartz$. The main estimate of this section is
given by
%
%
\begin{theorem}\label{theo:estglob} Assume $K$ is in $\CK_1$ or in
$\CK_0$ and let
$K_y=K+iy$ be as above. For every $\eps > 0$, there exist constants $\delta > 0$
and $C > 0$ such that for the norms defined by \eref{e:ab} one has
\begin{equ}[e:globBound]
\|u\|_{\delta,\delta} \le C\(\|u\|_{0,\eps} + \|K_y u\|\)
\end{equ}
holds for every $u \in \Schwartz$. The constants $C$ and $\delta $ are
independent of $y$ if $K\in\CK_1$.
\end{theorem}

Since $S^{\delta,\delta}$ is compactly embedded into $\L^2$, this
result implies:
%
%
\begin{corollary}
Let $K$ be as above. If there exist constants $\eps,C > 0$ such that
\begin{equ}[e:posBound]
\|u\|_{0,\eps} \le C\(\|u\| + \|K u\|\)\;,
\end{equ}
then $K$ has compact resolvent when considered as an operator acting on
$\L^2$.
\end{corollary}

\begin{proof}[of the Corollary]
Combining \eref{e:globBound} with \eref{e:posBound}, we get
\begin{equ}
\|u\|_{\delta,\delta} \le C\(\|u\| + \|K u\|\)\;.
\end{equ}
This implies that for $\lambda$ outside of the spectrum of $K$, the operator
$(K-\lambda)^{-1}$ is bounded from $\L^2$ into $S^{\delta,\delta}$. By
\lem{lem:est}(a), it is therefore compact.
\end{proof}

%
%
\begin{proof}[of \theo{theo:estglob}]
Let $\eps_*$ and $N_*$ be the values of the constants obtained in
estimate \eref{e:locest} of \theo{theo:loc}. 
Observe that \theo{theo:loc} also holds for any bigger value of $N_*$,
and we will assume $N_*$ is sufficiently large.

We choose $\eps >
0$. As a first
step, we will show that there exist constants $\delta$ and $C$ such that, for
any $x \in \R^n$ and $u \in \CC_0^\infty(\CB(x))$, the following estimate
holds:
\begin{equ}[e:estlocu]
\|u\|_{\delta,\delta} \le C\scalb^{-N_*} \|u\|_{\eps_*} + C
\scalb^{\eps/2} \|u\|\;.
\end{equ}
Denote by $J$ the smallest integer for which
\begin{equ}
J \ge 1 + {8N_* \over \eps}\;,
\end{equ}
and define
\begin{equ}[e:defdelta]
\delta = \min \Bigl\{2N_*, {\eps \over 2}, { \eps_* \over J} \Bigr\}\;.
\end{equ}
First, we note that when $A$ is a positive self-adjoint operator on some
Hilbert space $\CH$, one has the estimate
\begin{equ}[e:estA]
\|A u\|^{J} \le C\|A^{J} u\|\, \|u\|^{J - 1}\;,
\end{equ}
whenever both expressions make sense. In the case $J=2^j$ for $j$ an integer,
this can be seen by
a repeated application of the
Cauchy-Schwarz inequality. It was shown in \cite{KS59} to hold in the general
case as well.

We next use Jensen's inequality to write
\begin{equ}
\scalb^{N_* + \delta/2} \|\Lambda^\delta u\| \le C\({\|\Lambda^\delta u\| \over
\|u\|}\)^{J}\|u\| + C \scalb^{(N_* + \delta/2)\(1 + {1 \over J -1}\)}
\|u\|\;.
\end{equ}
Dividing this expression by $\scalb^{N_*}$ and using the definition of $J$, we
get
\begin{equs}
\scalb^{\delta/2} \|\Lambda^\delta u\| &\le C\scalb^{-N_*}\({\|\Lambda^\delta
u\| \over \|u\|}\)^{J}\|u\| \\
&\quad + C \scalb^{(N_* + \delta/2)\(1 +
{\eps/ (
8N_*)}\)-N_*} \|u\|\;.
\end{equs}
Using \eref{e:estA}, the fact that ${\eps \over 8N_*} \le {\eps - \delta \over 2N_*
+ \delta}$ by \eref{e:defdelta}, and $u\in
\CC_0^\infty(\CB(x))$, we get \eref{e:estlocu}.

In order to prove \theo{theo:estglob}, we use the following partition
of unity.
Let $\chi_0:\R\to[0,1]$ be a $\CC^\infty$ function with support in
$|x| < 1$ and satisfying $\sum_{i \in \Z} \chi_0(x-i) = 1$ for all $x\in \R$. The family
of functions
\begin{equ}
\CP = \{\chi_x:\R^n \to [0,1]\,|\,x\in \Z^n\}\;,
\end{equ}
defined by
\begin{equ}
\chi_x(z) = \prod_{j=1}^n \chi_0(z_j - x_j)\;,
\end{equ}
is therefore a partition of unity for $\R^n$. By construction, when
$x,x'\in\bf Z$ then $\chi_x$ and
$\chi_{x'}$ have disjoint support if there exists at least one index $j$ with
$|x_j^{} - x'_j|\ge 2$. We can therefore split $\CP$ into subsets $\left.\CP_k\right|_{k=1,\dots,3^n}$ such
that any two different functions belonging to the same $\CP_k$ have
disjoint supports. 

Consider next an arbitrary function $u \in \Schwartz$. We define $u_x =
\chi_x u$, and then the construction of the $\CP_k$ implies
\begin{equ}[e:fastfertig]
\sum_{x \in \Z^n} \|u_x\|_{0,\eps}\,\le\, 3^n \|u\|_{0,\eps}~.
\end{equ}
Using \eref{e:estlocu}, then \theo{theo:loc} and \eref{e:fastfertig},
we find
\begin{equs}
\|u\|_{\delta,\delta} &\le \sum_{x \in \Z^n} \|u_x\|_{\delta,\delta} \le
C\sum_{x \in \Z^n} \Bigl(\scalb^{-N_*} \|u_x\|_{\eps_*} + \scalb^{\eps/2}
\|u_x\|\Bigr) \\
&\le C\sum_{x \in \Z^n}\left (
\|u_x\|+\scalb^{-N_*} \|K_y u_x\|+\scalb^{\eps/2}
\|u_x\|
\right )\\
&\le C 3^n (\|u\|+\|u\|_{0,\eps}) + C\sum_{x \in \Z^n} \scalb^{-N_*}\|K_y u_x\|\;.
\end{equs}
For $k\in \{1,\ldots,3^n\}$ we now define
\begin{equ}
f_k \,=\, \sum_{\chi_{k,\ell} \in \CP_k} \scalb^{-N_*} \chi_{k,\ell}\;.
\end{equ}
With this notation, we have
\begin{equ}
\|u\|_{\delta,\delta} \le C \|u\|_{0,\eps} + C\sum_{k=1}^{3^n} \|K_y
f_k\,u\|\;.
\end{equ}
The claim \eref{e:globBound} thus follows if we can show that
\begin{equ}[e:estKf]
\|K_y f_k\,u\| \le C \|u\| + C \|K_y u\| \;.
\end{equ}
Since the $f_k$ are bounded functions, it suffices to estimate $\|[K,f_k] u\|$.
By construction, every derivative of $f_k$ decays like
$\scalb^{-N_*}$. 

Note that for sufficiently large $N_*$,
the functions $[X_j,f_k]$ and $\bigl[X_k,[X_j,f_k]\bigr]$ are bounded.
Since \theo{theo:loc} allows us to choose $N_*$ as large as we wish, 
\eref{e:estKf} follows from the estimate
$\|X_i\,u \|^2 \le \|u\| \,\|K_y u\|$.
\end{proof}
\subsection{Cusp}

%
%
Our statement about the cusp-like shape of the spectrum of $K$ is now
a consequence of Theorem \ref{theo:estglob}.
\begin{theorem}\label{theo:cusp}
Let $K \in \poly{2}{N}$ be of the type \eref{e:K}. Assume that the closure of $K$ in $\L^2$ is m-accretive and
that $K \in \CK_1$. Assume furthermore that there exist constants $\eps,C > 0$ such that
\begin{equ}[e:estapriori]
\|u\|_{0,\eps} \le C\(\|u\| + \|K_y u\|\)\;,
\end{equ}
for all $y \in \R$. Then, the spectrum of $K$ (as an operator on $\L^2$) is contained in the cusp
\begin{equ}
\{\lambda \in \C\;|\; \Re \lambda \ge 0\;,\; |\Im \lambda| \le C(1 + \Re
\lambda)^\nu\}\;,
\end{equ}
for some positive constants $C$ and $\nu$.
\end{theorem}
\begin{remark}In principle, our proofs give a constructive upper bound
on $\nu$. However, no attempt has been made to optimize this bound.
\end{remark}

\begin{proof}
The proof follows very closely that of Theorem~4.1 in \cite{HN02}, however we give 
the details for completeness. One ingredient we need is the following lemma:
\begin{lemma}\label{lem:Heinz}
Let $A : \L^2 \to \L^2$ be a maximal accretive operator that has $\Schwartz$ as a core.
Assume there exist constants $C,\alpha > 0$ for which
\begin{equ}
\|Au\| \le C\|u\|_{\alpha,\alpha}\;,\qquad \forall u\in\Schwartz\;.
\end{equ}
Then, for every $N \in \N$, there exists a constant $C_N$ such that
\begin{equ}
\|A^{1/N} u\| \le C_{N} \|u\|_{\alpha/N,\alpha/N}\;,\qquad \forall u\in S^{\alpha/N,\alpha/N}\;.
\end{equ}
\end{lemma}
\begin{proof}
By \lem{lem:est}(b), one can bound $\|u\|_{\alpha,\alpha}$ by
\begin{equ}
\|u\|_{\alpha,\alpha} \le C \bigl\|\bigl(\Lambda^{\alpha/2N}\bar \Lambda^{\alpha/N}\Lambda^{\alpha/2N}\bigr)^Nu\bigr\|\;.
\end{equ}
The generalized Heinz inequality presented in \cite{K61} then yields
\begin{equ}
\|A^{1/N} u\| \le C_{N}  \bigl\|\Lambda^{\alpha/2N}\bar \Lambda^{\alpha/N}\Lambda^{\alpha/2N}u\bigr\|\;.
\end{equ}
This concludes the proof of \lem{lem:Heinz}.
\end{proof}

We now turn to the proof of \theo{theo:cusp}. Since $K \in \poly{2}{N}$, one has for $\alpha = \max\{2,N\}$ the bound
\begin{equ}
\|(K+1)u\| \le C \|u\|_{\alpha,\alpha}\;,\quad \forall u \in \Schwartz\;.
\end{equ}
By \lem{lem:Heinz}, one can find for every $\delta>0$ an integer $M>0$ and a constant $C$ such that:
\begin{equ}[e:estKstarK]
\scal{u,\((K+1)^*(K+1)\)^{1/M}u} \le C \|u\|_{\delta,\delta}^2\;,
\end{equ}
Furthermore, \theo{theo:estglob} together with \eref{e:estapriori} yields constants $C$ and $\delta$ such that for
every $u\in\Schwartz$ and every $y\in\R$:
\begin{equ}[e:estunif]
\|u\|_{\delta,\delta}^2 \le C\(\|u\|^2 + \|(K+iy)u\|^2\)\;.
\end{equ}
Since $K$ is {\it m}-accretive by assumption, we can apply \cite[Prop.~B.1]{HN02} to get the estimate
\begin{equs}
{1 \over 4} |z+1|^{2/M}\|u\|^2 &\le \scal[b]{\((K+1)^*(K+1)\)^{1/M}u,u}
+ \|(K-z)u\|^2 \\
&\le C\|u\|_{\delta,\delta}^2 + \|(K-z)u\|^2 \;,
\end{equs}
where the second line is a consequence of \eref{e:estKstarK}.
Using \eref{e:estunif} and the triangle inequality for $z = \Re z + i\,\Im z$, we
get
\begin{equ}
{1 \over 4} |z+1|^{\eps /M}\|u\|^2 \le C\((1 + \Re z)^2\|u\|^2 +  \|(K-z)u\|^2
\)\;.
\end{equ}
Together with the compactness of the resolvent of $K$, this immediately implies that every $\lambda$ in the spectrum of $K$ satisfies the inequality
\begin{equ}
{1 \over 4} |\lambda+1|^{\eps /M}\|u\|^2 \le C(1 + \Re \lambda)^2\|u\|^2\;.
\end{equ}
This concludes the proof of \theo{theo:cusp}.
\end{proof}

\section{Examples}
\label{sect:example}

We present two examples in this section: A first, very simple one, and
a second which was the main motivation for this paper.
\subsection{Langevin equation for a simple anharmonic oscillator}
\label{sect:harm}

Our first example consists of one anharmonic oscillator which is in
contact with a stochastic heat bath at temperature $T$.  The
Hamiltonian of the oscillator is given by 
\begin{equ}
H(p,q) = {p^2 \over 2} + {\nu^2 q^2 \over 2} + \eps {q^4 \over 4}\;.
\end{equ}
For this model the associated spectral problem can be solved
explicitly when $\eps = 0$, because it is an harmonic oscillator. The
spectrum lies in a cone as shown in Fig.~\ref{fig:specharm}. We also
show that in first order perturbation theory in $\eps $, the
spectrum seems to 
form a non-trivial cusp, but this result remains conjectural, because
of non-uniformity of our bounds.

The Langevin equation for this system is
\begin{equ}[e:equosc]
dp = -\nu^2 q\,dt - \eps q^3\,dt - \gamma p\,dt + \sqrt{2\gamma T}\,dw(t)\;,\qquad dq = p\,dt\;,
\end{equ}
where $\gamma > 0$ measures the strength of the interaction between
the oscillator and the bath. Denote by $(\Omega,\Prob)$ the
probability space on which the Wiener process $w(t)$ is defined. We
write $\phi_{t,\omega}(x)$ with $\omega \in \Omega$ for the solution
at time $t$ for \eref{e:equosc} with initial condition $x = (p,q)$ and
realization $\omega$ of the white noise. The corresponding semigroups
acting on observables and on measures on $\R^2$ are given by 
\minilab{e:semigroup}
\begin{equs}
\(T_t f\)(x) &= \int_\Omega \(f\circ \phi_{t,\omega}(x)\)\,d\Prob(\omega)\;,\label{e:semobs}\\
\(T_t^* \mu\)(A) &= \int_\Omega \(\mu\circ \phi_{t,\omega}^{-1} (A)\)\,d\Prob(\omega)\;,\label{e:semmeas}
\end{equs}
where $A \subset \R^2$ is a Borel set. It is well-known that $$d\mu_T
(p,q) = \exp\(- H(p,q)/T\)\,dp\,dq$$ is the only stationary solution
for \eref{e:semmeas}.  

The It\^o formula yields for $f_t = T_t f$ the Fokker-Planck equation given by
\begin{equ}[e:FP]
\d_t f_t = \gamma T \d_p^2 f_t + p\,\d_q f_t - (\nu^2 q + \eps q^3 + \gamma p)\,\d_p f_t\;.
\end{equ}
We study \eref{e:FP} in the space $\CH_\beta = \L^2(\R^2,
d\mu_T)$. and make the change of variables $f_t = \exp\( H/(2T)\)F_t$ in
order to work in the unweighted space $\CH_0 = \L^2(\R^2, dp\,dq)$. Equation
\eref{e:FP} then becomes 
$\d_t F_t = - \tilde\CKK_\eps F_t$, where the differential operator $\tilde\CKK_\eps$ is given by
\begin{equ}
\tilde\CKK_\eps = - {\gamma T} \d_p^2  + {\gamma \over 4T} p^2 - {\gamma \over 2} - p\,\d_q + \nu^2 q\,\d_p + \eps q^3\,\d_p\;.
\end{equ}
By rescaling time, $p$ and $q$, one can bring $\tilde\CKK_\eps$ to the form
\begin{equ}
\CKK_\eps = {1 \over 2}\(-\d_p^2 + p^2 -1\) + \alpha(q\,\d_p - p\,\d_q)
+ c \eps q^3\,\d_p\;, 
\end{equ} 
where $\alpha = 2\sqrt{2T} \nu / \gamma$ and $c>0$.

The operator $K=\CKK_\eps$ is thus of the type \eref{e:K} with $X_0 =
\alpha(q\,\d_p - p\,\d_q) + c \eps q^3\,\d_p$ and $X_1 = \d_p$. 
We now verify the conditions of Section~\ref{sec:hypotheses}.
It is
obvious that these vector fields are of polynomial growth, thus
condition $a$ is satisfied. Since $[X_1,X_0] = -\alpha \d_q$, the
operator $\CKK_\eps$ satisfies condition $b_1$ as well, and so the
conclusion of \theo{theo:estglob} holds. 
Proceeding like in
\cite[Prop.~3.7]{EH}, one shows an estimate
of the type \eref{e:estapriori} (see also the proof of
\theo{theo:chain} below, where details are given). Therefore,
\theo{theo:cusp} applies, showing that the spectrum of $\CKK_\eps$ is
located in a cusp-shaped region. In fact, we show in the next
subsection that the cusp is a cone when $\eps =0$, and then we
study its perturbation to first order in $\eps $.
\subsubsection{First-order approximation of the spectrum of $\CKK_\eps$}

We will explicitly compute the spectrum and the corresponding eigenfunctions for $\CKK_0$ and then (formally) apply first-order perturbation theory to get an approximation to the spectrum of $\CKK_\eps$. We introduce the ``creation and annihilation'' operators
\begin{equ}
a = {p + \d_p \over \sqrt 2}\;, \quad a^* = {p - \d_p \over \sqrt 2}\;, \qquad
b = {q + \d_q \over \sqrt 2}\;, \quad b^* = {q - \d_q \over \sqrt 2}\;,
\end{equ}
in terms of which $\CKK_\eps$ can be written as 
\begin{equ}
\CKK_\eps = a^*a + \alpha( b^*a - a^*b) + c \eps q^3\,\d_p\;.
\end{equ}
With this notation, it is fairly easy to construct the spectrum of $\CKK_0$. Note first that $0$ is an eigenvalue for $\CKK$ with eigenfunction $\exp(-p^2/2 - q^2/2)$. This is actually the vacuum state for the two-dimensional harmonic oscillator in quantum mechanics (which is given by $a^*a + b^*b$), so we call this eigenfunction $\ket{\Omega}$.

\begin{figure}[h]
\begin{figurelist}{2}
\begin{center}
\mhpastefig[3/4]{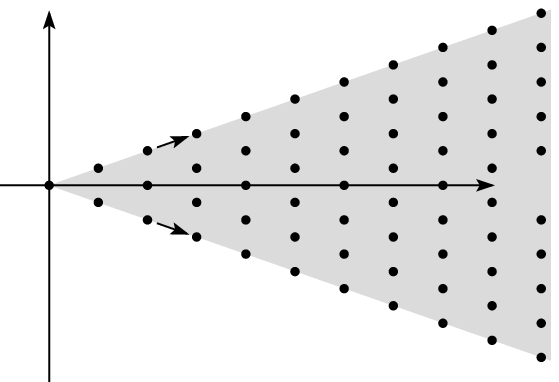}%
\caption{Spectrum of $\CKK_0$.}\label{fig:specharm}%
\end{center}%
&%
\begin{center}
\mhpastefig[3/4]{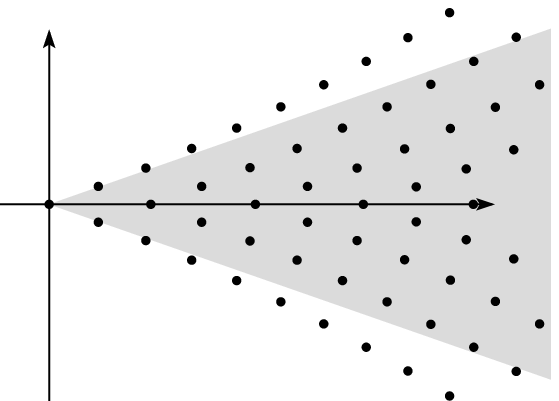}%
\caption{Approximate spectrum of $\CKK_\eps$.}\label{fig:pert}%
\end{center}
\end{figurelist}%
\vspace{-6mm}%
\end{figure}

A straightforward calculation shows that the creation operators $c_\pm^*$ defined by
\begin{equ}
c_\pm^* = a^* + \beta_\pm b^*\;,\qquad \beta_\pm = -{1 \over 2\alpha} \pm i {\sqrt{4\alpha^2-1} \over 2\alpha}\;,
\end{equ}
satisfy the following commutation relation with $\CKK_0$:
\begin{equ}
{}[\CKK_0, c_\pm^*] = \lambda_\pm c_\pm^*\;,\qquad \lambda_\pm = {1 \over 2} \pm i {\sqrt{4\alpha^2-1} \over 2} = -{\alpha \over \beta_\pm}\;.
\end{equ}
Therefore, $\lambda_0^{n,m} = n\lambda_+ + m\lambda_-$ with $n$ and $m$ positive integers are eigenvalues for $\CKK_0$ with eigenvectors given by
\begin{equ}
(c_+^*)^n(c_-^*)^m\ket{\Omega}\;.
\end{equ}
We conclude that for $\alpha > 1/2$ the spectrum of $\CKK_0$ consists of a triangular
grid located inside a cone (see Figure~\ref{fig:specharm}).  
\begin{remark}
Although the spectrum of $\CKK_0$ is located inside a sector,
$\CKK_0$ is {\em not} sectorial since the closure
of its
numerical range is the half-plane $\Re \lambda \ge 0$. 
\end{remark}

In order to do first-order perturbation theory for the spectrum of
$\CKK_\eps$  we also need the
eigenvectors for $\CKK_0^*$, which can be obtained by applying
successively $d_+^*$ and $d_-^*$ to $\ket{\Omega}$, where
\begin{equ}
d_\pm^* = a^* - \beta_\mp b^*\;.
\end{equ}
With this notation, $(d_+^*)^n(d_-^*)^m\ket{\Omega}$ is an eigenvector of $\CKK_0^*$ with eigenvalue $\bar \lambda_0^{n,m}$. 
By first-order perturbation theory, the eigenvalues of $\CKK_\eps$ are approximated by
\begin{equ}[e:deltanm]
\lambda_\eps^{n,m} \approx \lambda_0^{n,m} + c \eps \delta_{n,m} \;,\qquad \delta_{n,m} = {\scal{\Omega|d_-^m d_+^n q^3\d_p (c_+^*)^n(c_-^*)^m|\Omega} \over \scal{\Omega|d_-^m d_+^n(c_+^*)^n(c_-^*)^m|\Omega}}\;.
\end{equ}
The resulting spectrum\footnote{Actually the set $\{\lambda_0^{n,m} +
c \eps \delta_{n,m}~|~ n,m\ge0\}$.} is shown in Figure~\ref{fig:pert} (the sector
containing the spectrum of $\CKK_0$ is shown in light gray for
comparison). One clearly sees that the boundary of the sector bends to a
cusp. A (lengthy) explicit computation also shows that
\begin{equ}
\delta_{n,0} = -12n(n-1) {\bar\lambda_+ \over \sqrt{4\alpha^2-1}} + 9n{i\alpha \over \sqrt{4\alpha^2-1}}\;.
\end{equ}
In principle this confirms the cusp-like shape of the boundary, were
it not for the non-uniformity of the perturbation theory (in $n$).
\subsection{A model of heat conduction}

In this subsection, we apply our results to the physically more
interesting case of
a chain of nearest-neighbor interacting anharmonic oscillators coupled
to two heat baths at different temperatures. We model the chain by the
deterministic Hamiltonian system given by 
\begin{equ}
H = \sum_{i=0}^N \Bigl({p_i^2 \over 2} + V_1(q_i)\Bigr) + \sum_{i=1}^N V_2(q_i - q_{i-1})\;. 
\end{equ}
(We will give conditions on the potentials $V_1$ and $V_2$ later on.)
In order too keep notations short, we assume $p_i,q_i \in \R$, but one
could also take them in $\R^d$ instead. The two heat baths are modeled
by classical free field theories $\phi_L$ and $\phi_R$ with initial
conditions drawn randomly according to Gibbs measures at respective
inverse temperatures $\beta_L$ and $\beta_R$. (We refer to \cite{EPR}
for a more detailed description of the model.) It is shown in
\cite{EPR} that this model is equivalent to the following system of
stochastic differential equations: 
\begin{equs}[2]
	dq_i &= p_i\,dt\;,&i&=0,\ldots,N\;,\\[1mm]
	dp_0 &= -\V1'(q_0)\,dt + \V2'(\tilde q_1)\,dt + \rL\,dt\;, \\[1mm]
	dp_j &= -\V1'(q_j)\,dt - \V2'(\tilde q_{j})\,dt + \V2'(\tilde q_{j+1})\,dt\;,&j&=1,\ldots,N-1\;,
\\[1mm]
	dp_N &= -\V1'(q_N)\,dt - \V2'(\tilde q_{N})\,dt + \rR\,dt\;, \\[1mm]
	d\rL &= -\gamma_L \rL\,dt + \lambda_L^2\gamma_L q_0\,dt -
\lambda_L\sqrt{2\gamma_L T_L}\,dw_L(t)\;, \\[1mm]
	d\rR &= -\gamma_R \rR\,dt + \lambda_R^2\gamma_R q_N\,dt -
\lambda_R\sqrt{2\gamma_R T_R}\,dw_R(t)\;,
\end{equs}
where $T_i = \beta_i^{-1}$, $\gamma_i$ are positive constants describing the coupling of the chain to the heat baths, and $w_i$ are two independent Wiener processes. The variables $\rL$ and $\rR$ describe the internal state of the heat baths. If $T_L = T_R = T$, the equilibrium measure for this system is $d\mu_T (p,q,r) = \exp\(-G(p,q,r)/T\)\,dp\,dq\,dr$, where the ``energy'' $G$ is given by the expression
\begin{equ}
G(p,q,r) = H(p,q) + {\rL^2 \over 2\lambda_L^2} - q_0\rL + {\rR^2 \over 2\lambda_R^2} - q_N\rR  \;.
\end{equ}
If $T_L \neq T_R$, there is no way of guessing the invariant measure for the system. We can nevertheless make the construction of Section~\ref{sect:harm} with the reference measure $d\mu_{\tilde T}$ for some temperature
\begin{equ}
\tilde T > \max\{T_L,T_R\}\;,
\end{equ}
which is a stability condition, as one can see in \eref{e:stabf} below. The resulting operator $K=\CKK$ is given by
\begin{equ}[e:Kchain]
\CKK = X_L^*X_L^{} + X_R^*X_R^{} + f_L^2 + f_R^2 + X_0\;,
\end{equ}
where
\begin{equs}
X_{L,R} &= \lambda_{L,R}\sqrt{\gamma_{L,R} T_{L,R}} \d_{r_{L,R}}\;,\\
f_{L,R} &= \sqrt{\gamma_{L,R}(T_{L,R}/\tilde T -1)}(r_{L,R}-\lambda_{L,R}q_{0,N})\;, \label{e:stabf}\\
X_0 &= \nabla_{\!q} H \,\nabla_{\!p} - \nabla_{\!p} H \,\nabla_{\!q} + b_L(\rL-\lambda_L^2q_0)\d_{\rL} - \rL\d_{p_0}\\
&\quad + b_R(\rL-\lambda_R^2q_N)\d_{\rR} - \rL\d_{p_N}\;,
\end{equs}
with
\begin{equ}
b_{L,R} = {\gamma_{L,R} \over \lambda_{L,R}^2 \tilde T^2}(T_{L,R} - \tilde T)\;.
\end{equ}
We are now in a position to express the conditions of
Section~\ref{sec:hypotheses} in terms of sufficient conditions on the
potentials of the model. The first two assumptions guarantee that
$\CKK$ is in $\CK_1$.
%
%
\begin{assumption}\label{ass1}
There exist real numbers $n,m>0$ such that $D^\alpha V_1\in \poly{0}{2n-\alpha}$ and $D^\alpha V_2 \in \poly{0}{2m-\alpha}$ for $\alpha \le 2$.
\end{assumption}
%
%
%
%
\begin{assumption}\label{ass2}
There exists a constant $c>0$ such that $\V2''(x) > c$ for all $x\in\R$.
\end{assumption}
\begin{remark}
The second assumption states that there is a non-vanishing coupling between neighboring particles in every possible state of the chain.
\end{remark}
The verification that these assumptions imply $a$ is easy, and the
verification that $b_1$ holds can be found in
\cite{EPR,EH}. 

\begin{proposition}
Let $\CKK$ be defined as above and let $\V1$ and $\V2$ fulfill
Assumptions \ref{ass1} and \ref{ass2} above. Then $\CKK$ satisfies the
assumptions of \theo{theo:estglob} and satisfies Eq.\eref{e:globBound}
with $C$ and $\delta $ independent of $y$.
\end{proposition}
In order to show that the spectrum of $\CKK$ is located in a cusp-shaped region (\ie that the hypotheses of \theo{theo:cusp} hold), two more assumptions have to be made on the asymptotic behaviour of $\V1$ and $\V2$:
%
%
\begin{assumption}
The exponents $n$ and $m$ appearing in Assumption~\ref{ass1} satisfy $1 < n < m$.
\end{assumption}
\begin{remark}
The physical interpretation of the condition $n<m$ (actually $1\le
n\le m$ would probably work as well, see \cite{LTH}, but we could not
apply directly the results of \cite{EH}) goes as follows. If $n>m$,
the relative strength of the coupling between neighboring particles
decreases as the energy of the chain tends to infinity. Therefore, an
initial condition where all the energy of the chain is concentrated
into one single oscillator is ``metastable'' in the sense that the
energy gets transmitted only very slowly to the neighboring particles
and eventually to the heat baths. As a consequence, it is likely that
the convergence to a stationary state is no longer exponential in this
case, and so the operator $\CKK$ has probably not a compact resolvent
anymore. 
\end{remark}
Our last assumption states that the potentials and the resulting forces really grow asym\-pto\-ti\-cally like $|x|^n$ and $|x|^m$ respectively (and not just ``slower than'').
%
%
\begin{assumption}\label{ass4}
The potentials $\V1$ and $\V2$ satisfy the conditions
\begin{equs}[2]
\V1(x) &\ge c_1 \(1+\|x\|^2\)^n - c_2 \;,&\qquad x\V1'(x) &\ge c_3 \(1+\|x\|^2\)^n - c_4\;,\\
\V2(x) &\ge c_5 \(1+\|x\|^2\)^m - c_6 \;,&\qquad x\V2'(x) &\ge c_7 \(1+\|x\|^2\)^m - c_8\;,
\end{equs}
for all $x\in\R$ and for some positive constants $c_i$.
\end{assumption}
\begin{theorem}\label{theo:chain}
Let $\CKK$ be defined as above and let $\V1$ and $\V2$ fulfill assumptions \ref{ass1}--\ref{ass4} above. Then, $\CKK$ has compact resolvent and there exist positive constants $C$ and $N$ such that the spectrum of $\CKK$ is contained in the cusp
\begin{equ}
\Bigl\{\lambda \in \C\;\Big|\; \Re \lambda \ge 0\quad\text{and}\quad \Im \lambda \le C\(1 + |\Re \lambda|\)^N\Bigr\}\;.
\end{equ}
\end{theorem}
\begin{proof}
We will apply \theo{theo:cusp}, and need to check its assumptions.
It has been shown in \cite[Prop.~B.3]{EH} that $\CKK$ is {\it m}-accretive. The
fact that $\CKK \in \CK_1$ was checked above, and \eref{e:estapriori} was shown for $y=0$
in \cite[Prop.~3.7]{EH}. However, closer inspection of that proof
reveals that whenever $X_0$ was used, it only appeared inside a
commutator. Therefore, we can replace it by $X_0+iy$ without changing
the bounds. Thus, we have checked all the assumptions of
\theo{theo:cusp} and the proof of \theo{theo:chain} is complete.\end{proof}
\makeappendix{Proof of \lem{lem:est}}
\label{app:est}

The points {$a$}\ and {$b$} of \lem{lem:est} are standard results in
the theory of pseudodifferential operators (see \eg \cite[Vol.~III]{Ho} or,
more specifically, \cite{BCH,HTI,HTII}). The point {$c$}\ is an immediate
consequence of the Cauchy-Schwarz inequality combined with {$a$}. In order to
prove the points {$d$}\ and {$e$}, we first show the following intermediate
result:

%
%
\begin{lemma}\label{lem:bound}
Let $f:\R^n \to \R$ and $\alpha\in \R$. Let $k$ be the smallest {\em even}
integer such that $|\alpha| \le k$. Then, if $f$ satisfies
\begin{equ}
\sup_{y \in \R^n} |\d^\delta f(y)| < \kappa \;,\qquad \forall~~|\delta| \le k\;,
\end{equ}
the corresponding operator of multiplication is bounded from $S^{\alpha,\beta}$
into $S^{\alpha,\beta}$ and its operator norm is bounded by $C\kappa$. The
constant $C$ depends only on $\alpha$ and $\beta$.
\end{lemma}
\begin{proof}
By the definition of $S^{\alpha,\beta}$, it suffices to show that the
operator $\Lambda^\alpha f \Lambda^{-\alpha}$ is bounded by $C\kappa$
from $\L^2$ into $\L^2$. Since $f$ is obviously bounded by $\kappa$ as
a multiplication operator from $\L^2$ into $\L^2$, it actually suffices to
bound
$\Lambda^\alpha [f, \Lambda^{-\alpha}]$. Assume first that $\alpha \in
(0,2)$. In that case, we write
\begin{equ}
\Lambda^\alpha [f, \Lambda^{-\alpha}] = C_\alpha \int_0^\infty z^{-\alpha/2}
{\Lambda^\alpha \over z + \Lambda^2} [f,\Lambda^2]{1 \over z +
\Lambda^2}\,dz\;.
\end{equ}
The commutator appearing in this expression can be written as
\begin{equ}[e:commf]
{}[f,\Lambda^2] = \sum_{i=1}^n \(2 \d_i f\,\d_i + \d_i^2f\)\;.
\end{equ}
It is clear from basic Fourier analysis that $\|\d_i (z +
\Lambda^2)^{-1/2}\| \le 1$ and therefore
\begin{equ}
\|[f,\Lambda^2](z + \Lambda^2)^{-1/2}\| \le C\kappa\;.
\end{equ}
Furthermore, the spectral theorem tells us that for any function $F$,
$\|F(\Lambda^2)\|$ is bounded by $\sup_{\lambda \ge 1} F(\lambda)$. Therefore
there exists a constant $C$ independent of $z>0$ such that
\begin{equ}
\|\Lambda^\alpha (z + \Lambda^2)^{-1}\| \le {C \over 1 + z^{1-\alpha/2}}\;.
\end{equ}
Combining these estimates shows the claim when $\alpha \in
(0,2)$. The case $\alpha = 2$ follows from the boundedness of
$[f,\Lambda^2]\Lambda^{-2}$. Values of $\alpha$ greater than $2$ can be
obtained by iterating the relation
\begin{equ}
\Lambda^{\alpha+2}f \Lambda^{-\alpha -2} = \Lambda^\alpha f \Lambda^{-\alpha} +
\Lambda^\alpha [f,\Lambda^2] \Lambda^{-\alpha-2}\;.
\end{equ}
Using \eref{e:commf}, the fact that $\d_i$ commutes with $\Lambda$,
and the fact that $\d_i \Lambda^{-2}$ is bounded, we can reduce this to the
previous case, but with two more derivatives to control. The
case $\alpha < 0$ follows by considering adjoints. This
concludes the proof of \lem{lem:bound}.
\end{proof}

\begin{remark}
Since the direct and the inverse Fourier transforms both map
$S^{\alpha,\beta}$ continuously into $S^{\beta,\alpha}$, the above
lemma also holds for bounded functions of $\d_y$ and not only for
bounded functions of $y$.
\end{remark}

We are now ready to turn to the

%
%
{\noindent\textbf{Proof of point \textit{d}.}}
Let $X\in\poly{k}{N}$.
We first consider $\gamma \in
(-2,0)$. Since, in Fourier space, $\Lambda^2$ is a multiplication operator by a
real positive function, we can write
\begin{equ}{}
[X,\Lambda^\gamma] = C_\gamma \int_0^\infty z^{\gamma/2} {1 \over z +
\Lambda^2} [X,\Lambda^2] {dz \over z + \Lambda^2}\;.
\end{equ}
In order to bound this expression, we define $B = [X,\Lambda^2]$,
commute $B$ with the resolvent, and obtain
\begin{equs}{}
[X,\Lambda^\gamma] &= C_\gamma \int_0^\infty z^{\gamma/2} {\Lambda^{2-\gamma}
\,dz \over (z + \Lambda^2)^2}  \, \Lambda^{\gamma-2}B + C_\gamma \int_0^\infty
 {z^{\gamma/2} \over (z + \Lambda^2)^2}[B,\Lambda^2] {dz \over z +
\Lambda^2}\,.\\
\end{equs}
The first term equals $C'_\gamma \Lambda^{\gamma-2}B$ because $\int_0^\infty z^{\gamma/2} x^{2-\gamma} (z +
x^2)^{-2}\,dz$ does not depend on $x>0$. This, in turn, is bounded from
$S^{\alpha,\beta}$ into $S^{\alpha+1-k-\gamma,\beta-N}$ using
$B\in\poly{k+1}{N}$ and \lem{lem:est}(b).
To bound the second term, we rewrite
$$
\int_0^\infty
 {z^{\gamma/2} \over (z + \Lambda^2)^2}[B,\Lambda^2] {dz \over z +
\Lambda^2}\,=\,
\int_0^\infty
 {z^{\gamma/2} \Lambda ^{1-\gamma}\over (z + \Lambda^2)^2}\cdot\Lambda
 ^{\gamma-1}[B,\Lambda^2]\Lambda ^{-2}\cdot {\Lambda ^2 \over z + 
\Lambda^2}\,dz~.
$$
The factor $\Lambda^2(z + \Lambda^2)^{-1}$
is bounded from $S^{\alpha,\beta}$ into itself, uniformly
in $z$.
Using \lem{lem:est}(b) as before, we see that the factor
$\Lambda^{\gamma-1}[B,\Lambda ^2]\Lambda^{-2}$ is bounded from $S^{\alpha,\beta}$ into
$S^{\alpha+1-k-\gamma,\beta-N}\equiv S^{\alpha ',\beta '}$. 
Finally, using \lem{lem:bound} and counting powers, we see that the
first factor has norm bounded by $\CO(z^{-3/2})$ for large $z$ and
$\CO(z^{\gamma /2})$ for $z$ near $0$ as a map from $S^{\alpha ',\beta
'}$ to itself. This proves the first statement of \lem{lem:est}(d).
The second one is proven similarly and is left to the reader.

{\noindent\textbf{Proof of point \textit{e}.}} Recall that we want to bound
\begin{equ}
I = |\scal{f,Xg}_{\alpha,\beta} - \scal{X^T f,g}_{\alpha,\beta}|\;,
\end{equ}
where $X \in \poly{k}{N}$ and $X^T$ denotes the formal adjoint
(in $\L^2$) of $X$. We write this as
\begin{equ}
I =
\scal{[\weight^{-2\beta}\Lambda^{-2\alpha},X^T]\Lambda^{2\alpha}
\weight^{2\beta}f,g}_{\alpha,\beta}\;.
\end{equ}
We rewrite the operator as
\begin{equ}{}
[\weight^{-2\beta}\Lambda^{-2\alpha},X^T]\Lambda^{2\alpha}\weight^{2\beta} =
\weight^{-2\beta}[\Lambda^{-2\alpha},X^T]\Lambda^{2\alpha}\weight^{2\beta} +
[\weight^{-2\beta},X^T]\weight^{2\beta}\;.
\end{equ}
The second term is in $\poly{k-1}{N}$ by inspection, and the required
bound follows at once from
\lem{lem:est}(b,c).
The 
first term is bounded similarly by using \lem{lem:est}(d,b,c). This concludes the proof
of \lem{lem:est}.\qed

\bibliographystyle{Martin}
\markboth{\sc \refname}{\sc \refname}
\bibliography{refs}
\end{document}